\documentclass{PoS}
\usepackage{cite}
\usepackage{amssymb}
\usepackage{amsmath}
\usepackage{latexsym}
\usepackage{multirow}
\usepackage{xspace}
\usepackage{array}
\usepackage{url}
\usepackage{bbding}
\usepackage{afterpage}

\newcommand{\SqrtS}{\sqrt{S}}

\newcommand{\TB}{\ensuremath{\tan\beta}\xspace}
\newcommand{\bbar}{\ensuremath{b\bar{b}}\xspace}
\newcommand{\mbeff}{\ensuremath{m_b^{\text{eff}}}\xspace}

\newcommand{\stau}{\ensuremath{{\tilde{\tau}}_{1}}\xspace}
\newcommand{\staui}{\ensuremath{{\tilde{\tau}}_{i}}\xspace}
\newcommand{\stauj}{\ensuremath{{\tilde{\tau}}_{j}}\xspace}

\newcommand{\supertau}{\ensuremath{\tilde\tau}\xspace}
\newcommand{\stautwo}{\ensuremath{{\tilde{\tau}}_{2}}\xspace}
\newcommand{\mstau}{\ensuremath{m_{\stau}}\xspace}

\newcommand{\HH}{\ensuremath{H^{0}}\xspace}
\newcommand{\mHH}{\ensuremath{m_{H^{0}}}\xspace}
\newcommand{\hh}{\ensuremath{h^{0}}\xspace}

\newcommand{\staubar}{\ensuremath{{{\tilde{\tau}}}_1^{*}}\xspace}
\newcommand{\staustaubar}{\ensuremath{{{\tilde\tau}}_{1}^{\phantom{*}}\staubar}\xspace}

\newcommand{\Hhiggs}{\ensuremath{{H}^0}\xspace}

\newcommand{\thetastau}{\ensuremath{\theta_{{\tilde{\tau}}}}\xspace}

\newcommand{\Atau}{\ensuremath{A_{\tau}}\xspace}
\newcommand{\Atop}{\ensuremath{A_{\mathrm{t}}}}
\newcommand{\Abottom}{\ensuremath{A_{\mathrm{b}}}}
\newcommand{\stauR}{\ensuremath{{\tilde{\tau}_{\mathrm{R}}}}\xspace}
\newcommand{\stauL}{\ensuremath{{\tilde{\tau}_{\mathrm{L}}}}\xspace}

\newcommand{\fbarn}{\mathrm{fb}}

\newcommand{\squark}{\tilde{q}}

\newcommand{\neu}{\ensuremath{\tilde{\chi}^{0}}}

\newcommand{\MeV}{{\rm Me\kern -1pt V}}
\newcommand{\GeV}{{\rm Ge\kern -1pt V}}
\newcommand{\TeV}{{\rm Te\kern -1pt V}}

\newcommand{\DY}{Drell-Yan\xspace}

\newcommand{\mhalf}{\ensuremath{m_{1/2}}\xspace}
\newcommand{\mzero}{\ensuremath{m_{0}}\xspace}

\newcommand{\ord}{{\cal O}}

\newcommand{\be}{\begin{eqnarray*}}
\newcommand{\ee}{\end{eqnarray*}}
\newcommand{\bee}{\begin{eqnarray}}
\newcommand{\eee}{\end{eqnarray}}
\renewcommand{\eqref}[1]{eq.~(\ref{#1})}

\newcommand{\figref}[1]{figure~\ref{#1}}

\newcommand{\figsref}[2]{figures~\ref{#1} and~\ref{#2}}
\newcommand{\eg}[0]{\textit{e.g.}\xspace}
\newcommand{\cf}[0]{\textit{cf.}\xspace}
\newcommand{\ie}[0]{\textit{i.e.}\xspace}

\newcommand{\Prospino}[0]{\texttt{Prospino~2}\xspace}
\newcommand{\FeynArts}[0]{\texttt{FeynArts~3.6}\xspace}
\newcommand{\FormCalc}[0]{\texttt{FormCalc~7.0}\xspace}
\newcommand{\LoopTools}[0]{\texttt{LoopTools~2.6}\xspace}
\newcommand{\FeynHiggs}[0]{\texttt{FeynHiggs~2.7.4}\xspace}

\newcommand{\pT}{\ensuremath{p^\mathrm{T}}\xspace}

\title{Direct stau production at the LHC}

\ShortTitle{Direct stau production at the LHC}

\author{\speaker{Jonas M. Lindert}\\
        Max-Planck-Institut f\"ur Physik, 
 F\"ohringer Ring 6, 
 D-80805 M\"unchen, Germany\\
       E-mail: \email{lindert@mpp.mpg.de}}

\author{Frank D. Steffen\\
         Max-Planck-Institut f\"ur Physik, 
 F\"ohringer Ring 6, 
 D-80805 M\"unchen, Germany\\
       E-mail: \email{steffen@mpp.mpg.de}}

\author{Maike K. Trenkel\\
        Phenomenology Institute, Department of Physics, 
 University of Wisconsin-Madison, 
 1150 University Avenue, 
 Madison, Wisconsin 53706, USA\\
       E-mail: \email{trenkel@hep.wisc.edu}}

\abstract{
We investigate the direct production of supersymmetric scalar taus at the LHC.  
We present the general calculation of the dominant cross section contributions  
for hadronic stau pair production within the MSSM, 
taking into account left-right mixing of the stau eigenstates. 
We find that $b$-quark annihilation and gluon fusion 
can enhance the cross sections by more than one order of magnitude 
with respect to the Drell-Yan predictions.
For long-lived staus, we consider CMSSM parameter regions with such enhanced cross sections
and possible consequences from recent searches. We find that regions of exceptionally
small stau yields, favoured by cosmology, are in tension with a recent 
CMS limit on $m_{\stau}$.
}

\dedicated{MPP-2012-71}

\FullConference{Proceedings of the Corfu Summer Institute 2011 "School and Workshops on Elementary Particle Physics and Gravity"\\
		 September 4-18, 2011\\
		 Corfu, Greece}

\begin{document}

% ____________________________________________________________________
\section{Introduction}
\label{sec:intro}
% ____________________________________________________________________

The ongoing experiments at the Large Hadron Collider (LHC) set limits for
new physics at the $\TeV$ scale in an unprecedented way. Amongst the most 
promising candidates for new physics are supersymmetric (SUSY) theories.
Within this class of theories, scenarios with a scalar tau \supertau being the lightest 
ordinary SUSY particle (LOSP) are often considered. In R-parity-conserving scenarios 
the lightest SUSY particle (LSP) is then usually assumed to be extremely weakly interacting, 
\eg, a gravitino or an axino. Such scenarios offer spectacular
phenomenological signatures at colliders: long-lived charged massive particles (CHAMPs). 
Furthermore, they can be associated with intriguing early Universe cosmology, \cf
\cite{Steffen:2008qp} and references therein. For example, an overly abundant charged relic can spoil the 
successful predictions of BBN. Resulting strong bounds can be evaded in parameter 
regions where the stau annihilates efficiently, \eg, via a heavy Higgs resonance or due 
to large stau--higgs couplings \cite{Ratz:2008qh,Pradler:2008qc}. In scenarios with broken 
R-parity, the \supertau can also be the LSP that decays promptly (or delayed in 
scenarios with partly broken R-parity) into Standard Model (SM) particles.

In this work we present theoretical predictions for direct \staustaubar production at the LHC
within the minimal SUSY extension of the Standard Model (MSSM) independent of a possible 
longevity of the stau. We include Drell-Yan processes as well as $b$-quark annhilation and 
gluon fusion. Special attention is given to mixing effects due to large Yukawa couplings. After 
evaluating numerical results in the phenomenological MSSM to investigate possible enhancement 
effects, we interpret our results in the CMSSM assuming a long-lived \stau.  
This proceeding is mainly based on Ref. \cite{staupaper} where also more references can be found. 
One new aspect which has not been discussed in \cite{staupaper} is the interpretation of recent experimental 
exclusion limits set by the CMS experiment at the LHC.

% ____________________________________________________________________
\section{Direct production of stau pairs at the LHC}
\label{sec:production}
\label{sec:stauprod}
% ____________________________________________________________________
Within the MSSM, stau pairs can be produced directly at the LHC,
\begin{align}
pp \to \staui^{}\stauj^*,
\end{align}
where $\tilde{\tau}_{i,j}$ denotes any of the two stau mass eigenstates. 
Due to off-diagonal elements in the stau mass matrix large mixing 
between gauge eigenstates  $(\stauL, \stauR)$ can occur for the mass 
eigenstates $(\stau, \stautwo)$ and may not be neglected for third generation
sleptons. This mixing can become large and is proportional to the SUSY 
parameters $\mu$, $\tan\beta$ and \Atau.
Here we concentrate on the direct production of the lighter 
$\staustaubar$ pairs, particularly in parameter regions with large mixing. 
Production rates of ${\tilde\tau}_2^{\phantom{*}} {\tilde\tau_2}^*$ and similarly 
${\tilde\tau_1}^{\phantom{*}} {\tilde\tau_2}^*$ can be obtained in close
analogy. \\

At the LHC the leading contribution to direct $\staustaubar$ production 
up to order $\ord(\alpha_s^2 \alpha^2)$ are given by the following channels:
\begin{itemize}
\item$q\bar{q}$ induced Drell-Yan type processes at $\ord(\alpha^2)$, see
\figref{fig:feynman_dybb}\,(a),  and corresponding NLO (SUSY)QCD corrections at 
$\ord(\alpha_s^2 \alpha^2)$. 

\item \bbar annihilation, mediated by the neutral gauge
bosons ($\gamma$, $Z$) and by the neutral CP-even Higgs bosons ($h^0$, $H^0$) at $\ord(\alpha^2)$
(however suppressed by the low bottom-quark PDFs), shown in \figref{fig:feynman_dybb}\,(b).

\item gluon-gluon fusion at order $\ord(\alpha_s^2 \alpha^2)$, mediated
by a quark or squark loop, as shown in \figref{fig:feynman_gluglu}.

\end{itemize}

The \DY~production cross section at leading order depends only on the stau mass $\mstau$ 
and the stau mixing angle $\thetastau$. The parametric dependence of the NLO corrections
on the SUSY parameters in general is small.  
Although suppressed by the low bottom-quark density inside protons the \bbar 
channel can be enhanced by on-shell Higgs propagators and by the bottom-Higgs 
and the stau-Higgs couplings in certain regions of the SUSY parameter space. In the basis
of mass eigenstates, the stau-Higgs coupling is proportional to $\sin\thetastau$ and 
thus becomes important for large mixing in the stau sector. Both the stau-Higgs and the 
bottom-Higgs couplings are proportional to $\mu$ and the respective trilinear coupling $A_{\tau / b}$. 
In order to study possible enhancement effects in direct $\staustaubar$ production due to large 
stau--Higgs couplings, parameter regions with relatively large \TB are to be considered. 
Here radiative corrections to the $\bbar h^0/\bbar H^0$ vertex can be
important and drive down the cross section compared to the tree-level result. 
Leading \TB-enhanced corrections can be resummed to all orders in pertubation
theory by using an appropriate effective bottom-quark mass, \mbeff, and effective
$\bbar h^0/\bbar H^0$ couplings. Here we adopt this approach, as explained in detail in
appendix B of \cite{staupaper}.
%%%%%%%%%%%
%
Finally, even though gluon-induced contributions are formally of higher orders, $\ord(\alpha_s^2
\alpha^2)$, they can give sizeable contributions at the $pp$-machine LHC at high
center-of-mass energies where the $gg$ luminosity is significantly higher than the
$q\bar{q}$~luminosity. As the \bbar channel, the gluon-gluon channel depends strongly
on the parameters in the stau-Higgs sector. 

% % % % % % % % % % % % % % % % % % % % % % % % 
\FIGURE[t]{
\includegraphics{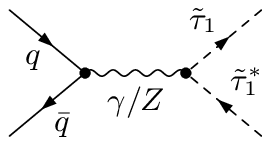}
\hspace*{2.5cm}
\includegraphics{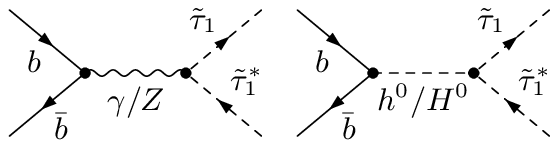}\\[1ex]
\small (a) \hspace*{6cm} (b) \hspace*{2cm}
\caption{Feynman diagrams for stau pair production (a) via the Drell-Yan process and (b) via \bbar annihilation. Here $q= u,d,c,s$.}
\label{fig:feynman_dybb}
}
% % % % % % % % % % % % % % % % % % % % % % % % 
\FIGURE[t]{
\includegraphics{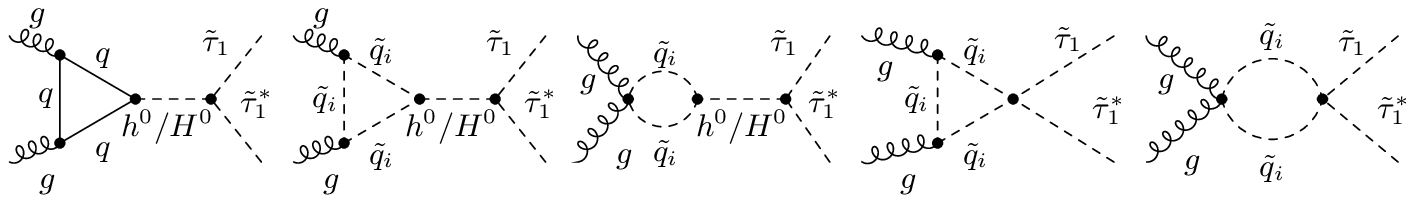}
\caption{Feynman diagrams for the gluon fusion contribution to stau pair production. 
The quarks~$q$ and squarks $\squark_i$, $i=1,2$, running in the loops can be of any flavor.}
\label{fig:feynman_gluglu}
}
% % % % % % % % % % % % % % % % % % % % % % % % 

We use the programs \FeynArts and \FormCalc with \LoopTools to generate and 
calculate the amplitudes corresponding to the Feynman diagrams of~\figsref{fig:feynman_dybb}{fig:feynman_gluglu}. 
The Higgs boson masses and the $H^0$ width are computed with
\FeynHiggs.
The QCD and SUSY-QCD corrections at NLO for the Drell-Yan channel calculated in \cite{Beenakker:1999xh} we evaluate with \Prospino and scale our cross sections with the resulting $K$-factors.
Furthermore, we use a resummed effective $\bbar h^0/\bbar H^0$ 
vertex for the gluon fusion and \bbar contributions, as explained above.
Since we do not include (expected to be positive) higher-order QCD and SUSY-QCD 
corrections to the Higgs-mediated channels, our analysis gives a conservative estimate 
of the enhancement effects from the \bbar-annhilation and gluon fusion production channels.

At the LHC, the gluon-fusion and \bbar-annhilation processes 
with an $s$-channel Higgs boson can become resonant in regions of the 
SUSY parameter space in which the Higgs boson $H^0$ is heavier than the two produced staus. 
In parameter regions with $m_{H^0} \ge 2 \mstau$
we therefore include the total decay width of the $\HH$ boson, $\Gamma_\HH$ in the propagator,
\be
\label{eq:higgswidth}
\frac{1}{p^2-\mHH^2} \longrightarrow
\frac{1}{p^2-\mHH^2+i\mHH\Gamma_{H^0}} \, .
\ee

% ____________________________________________________________________
\section{Numerical results}
\label{sec:results}
% ____________________________________________________________________

The cross section for direct stau production depends mainly on  $\mstau$, $m_{H^0}$, 
$\TB$, and on $\thetastau$ (or equivalently on $\mu$ and $A_{\tau}$). Let us now 
investigate the dependence on these parameters.

As a starting point, we choose a $\stau$-LOSP scenario with 
moderate squark masses and a large stau--Higgs coupling,
fixed by the following soft-breaking parameters at the low scale:
\begin{align}
\begin{split}
M_1 &= M_2 = M_3 = 1.2~\text{TeV}, \qquad
\Atop=\Abottom=\Atau=600~\text{GeV},
\\
m_{\tilde Q_i} &= m_{\tilde U_i} = m_{\tilde D_i} = 1~\text{TeV}, \qquad
m_{\tilde L_{1/2}} = m_{\tilde E_{1/2}} =  500~\text{GeV},
\label{eq_SUSYinputs}
\end{split}
\end{align}
If not otherwise stated, we choose, 
\begin{align}
\begin{split}
\thetastau &= 45^{\circ}, \qquad 
\mstau = 200~\text{GeV}, \qquad
\\ 
\TB &= 30, \qquad~
\mu = 500~\text{GeV}, \qquad
m_A = 400~\text{GeV},
\end{split}
\label{eq_STAUinputs}
\end{align}
as inputs for the third-generation sleptons and the Higgs sector. 

From these input parameters, we calculate the physical MSSM parameters 
using tree-level relations for sfermions, neutralinos, and charginos.  
Physical masses are then passed to \Prospino to calculate 
the Drell-Yan K--factors at NLO in QCD and SUSY-QCD. The NLO
corrections to the Drell-Yan channel typically amount to
$20-40~\%$ in the considered parameter space.

% % % % % % % % % % % % % % % % % % % % % % % % 
\FIGURE[t]{
\includegraphics[width=.49\textwidth,]{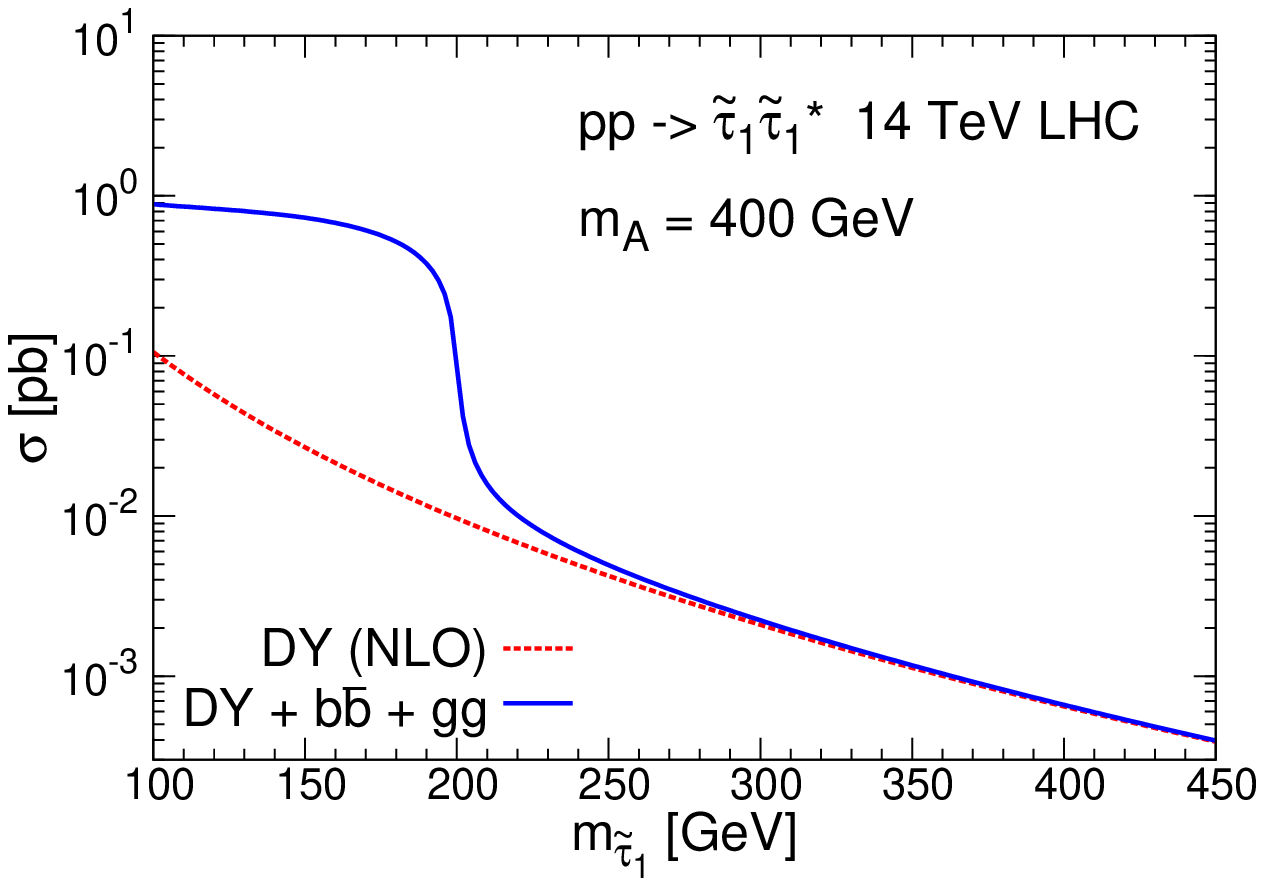}
\includegraphics[width=.49\textwidth,]{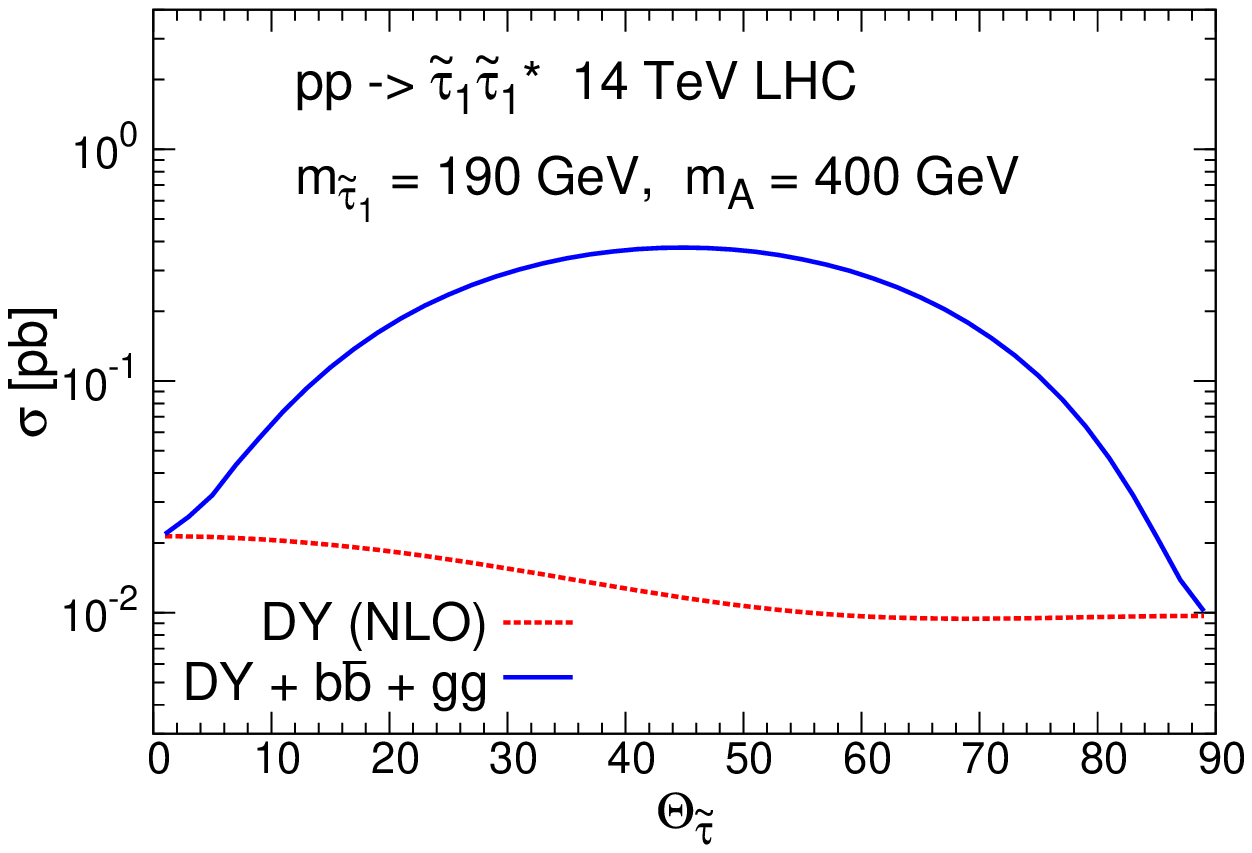}\\[.5ex]
{\small \hspace*{2cm} (a) \hspace*{.47\linewidth} (b)} \\[2.5ex]
\includegraphics[width=.49\textwidth,]{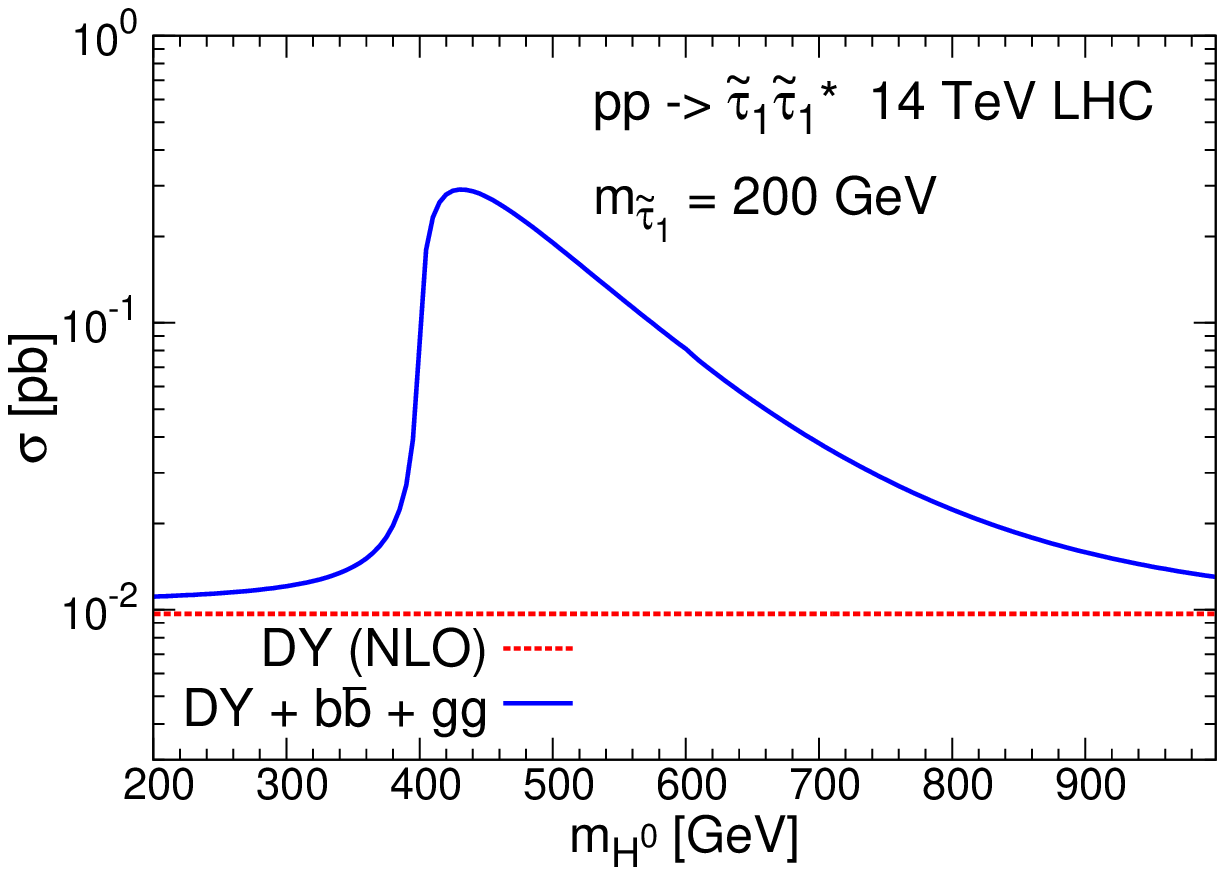}
\includegraphics[width=.49\textwidth,]{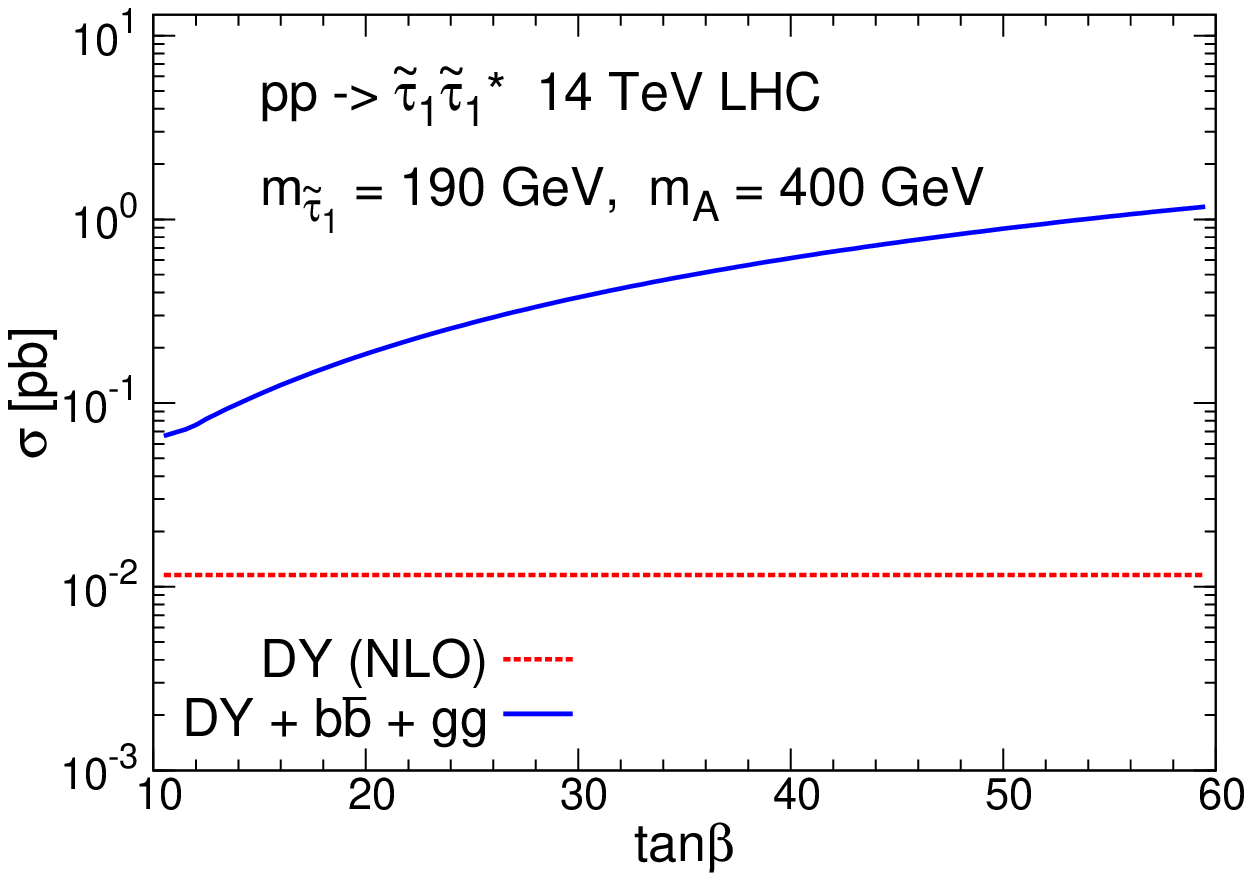}\\[.5ex]
{\small \hspace*{2cm} (c) \hspace*{.47\linewidth} (d)} \\[-1ex] 
 % scan_mtau1_tb30_max_channels.ps: 504x720 pixel, 72dpi, 17.78x25.40 cm, bb=0 0 504 720
\caption{Cross section of direct \stau production at the LHC with $\SqrtS =
14~\TeV$ as a function of (a)~\mstau, (b)~\thetastau, (c)~\mHH,
and (d)~\TB. No kinematical cuts are applied. 
SUSY input parameters as given in (\ref{eq_SUSYinputs}) and (\ref{eq_STAUinputs}).
Note that here $\mHH \approx m_A$ (decoupling limit). 
}
\label{fig:crosssections1}
}
% % % % % % % % % % % % % % % % % % % % % % % % 

% % % % % % % % % % % % % % % % % % % % % % % % 
\FIGURE[t]{
\includegraphics[width=.49\textwidth,]{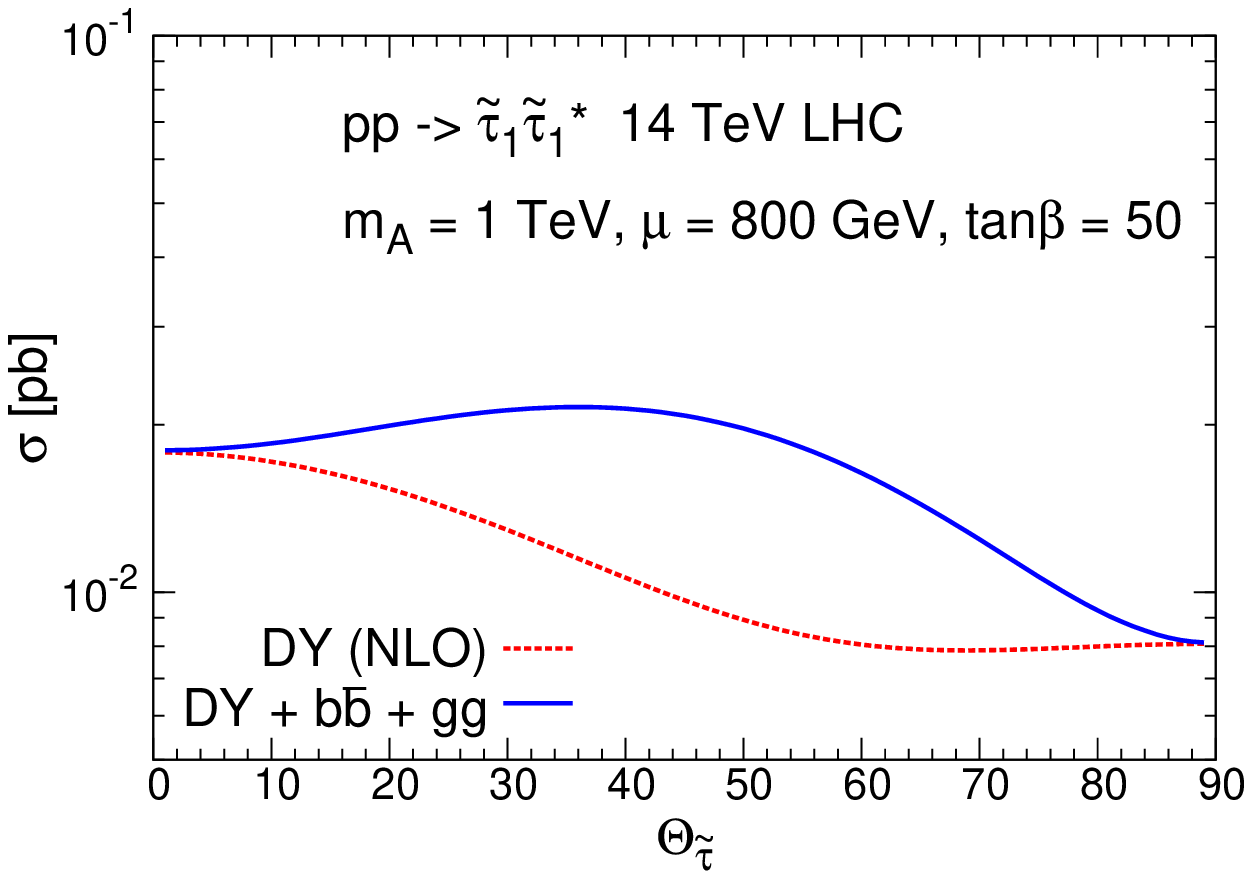}
\includegraphics[width=.49\textwidth,]{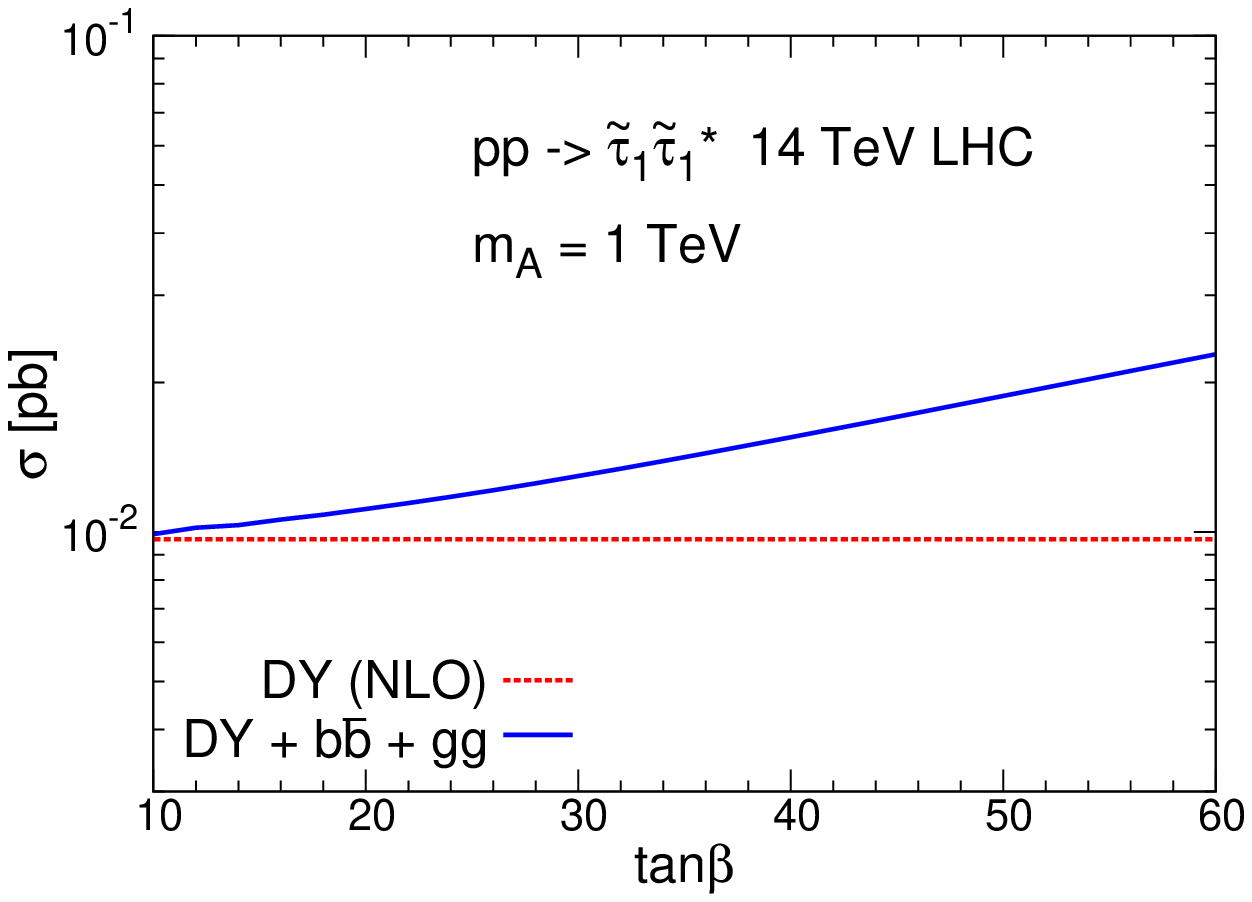}\\[.5ex]
{\small \hspace*{2cm} (a) \hspace*{.47\linewidth} (b)} \\[-1ex] 
 % scan_mtau1_tb30_max_channels.ps: 504x720 pixel, 72dpi, 17.78x25.40 cm, bb=0 0 504 720
\caption{Cross section of direct \stau production at the LHC with $\SqrtS =
14~\TeV$ as a function of (a)~\thetastau and (b)~\TB. No kinematical cuts are applied. 
SUSY input parameters as given in (\ref{eq_SUSYinputs}) and (\ref{eq_STAUinputs}) if not stated otherwise.
Note that here $\mHH \approx m_A = 1~\TeV$ (decoupling limit). 
}
\label{fig:crosssections2}
}
% % % % % % % % % % % % % % % % % % % % % % % % 

In \figref{fig:crosssections1} we show the direct production cross section for
\stau-pairs at the LHC with $\SqrtS = 14~\TeV$ as a function of (a)~\mstau, (b)~\thetastau, 
(c)~\mHH, and (d)~\TB. 
In figures \ref{fig:crosssections1}\,(b) and (d), we move to $\mstau=190~\GeV$ where stau
production is possible via an on-shell \HH.
The dashed (red) lines show the Drell-Yan (DY) cross section at NLO, 
whereas the solid (blue) lines include the additional $\bbar$ and $gg$~contributions. 
The Drell-Yan cross section depends on \mstau and \thetastau only. It decreases strongly
for increasing $\stau$ masses and varies roughly by a factor slightly larger than $2$ with
$\thetastau$, as shown in \figref{fig:crosssections1}\,(b), being largest for $\thetastau
\approx 0$, \ie, an almost left-handed \stau.
The impact of the $\bbar$ and $gg$ channels depends strongly 
on the mass hierarchy between \stau and \Hhiggs,
as can clearly be seen in Figures~\ref{fig:crosssections1}\,(a) and~(c).
If $\mHH > 2 \mstau$, these additional channels 
can change the direct production cross section by more than one order of magnitude
with respect to the Drell-Yan result.
At the threshold $\mHH = 2 \mstau$, the $\bbar$ and $gg$ contributions drop steeply 
and are only marginally important for $\mHH\ll 2 \mstau$.

Figures~\ref{fig:crosssections1}\,(b) and (d)
illustrate the dependence of the total direct production cross section
on the parameters $\thetastau$ and $\TB$ that govern the
stau-Higgs-coupling strength.
The additional contributions from the $\bbar$ and $gg$
channels are tiny in cases of very small mixing, $\thetastau\to 0, \pi$, 
but become most important for maximal mixing, \ie, at $\thetastau\approx \pi/4$.
For very large $\tan\beta$, additional contributions push up the total direct production cross
section by up to two orders of magnitude and are still sizeable for small \TB.

Let us now turn to a scenario where the \HH is very heavy and thus almost decoupled, $\mHH
= 1~\TeV$. We again investigate the dependence of the total cross section on \thetastau
and \TB, shown in \figref{fig:crosssections2}, where we focus on enhanced
\stau-\staubar-\hh couplings. In \figref{fig:crosssections2}\,(a) we choose
$\mu=800~\GeV$ and $\TB=50$ to increase this coupling. All other parameters are fixed
according to (\ref{eq_SUSYinputs}) and (\ref{eq_STAUinputs}).
Again, contributions from additional $\bbar$ and $gg$ channels can be sizeable and the
enhancement amounts to a factor of two or three when considering very large values of \TB
and maximal mixing $\thetastau \approx \pi/4$. However, here dominant contributions come
mainly from off-shell \hh diagrams together with large couplings. Thus, the relative
importance between the $gg$ channel and the $\bbar$ channel can be different compared
to on-shell \HH production, as the two Higgses couple differently to the squark
loops. 
We want to note that, despite the large couplings, all 
considered parameter points are in agreement with bounds from the potential 
occurrence of (color and) charge breaking (CCB) minima \cite{Hisano:2010re}.

% % % % % % % % % % % % % % % % % % % % % % % % 
\FIGURE[t]{
\includegraphics[width=.75\textwidth,]{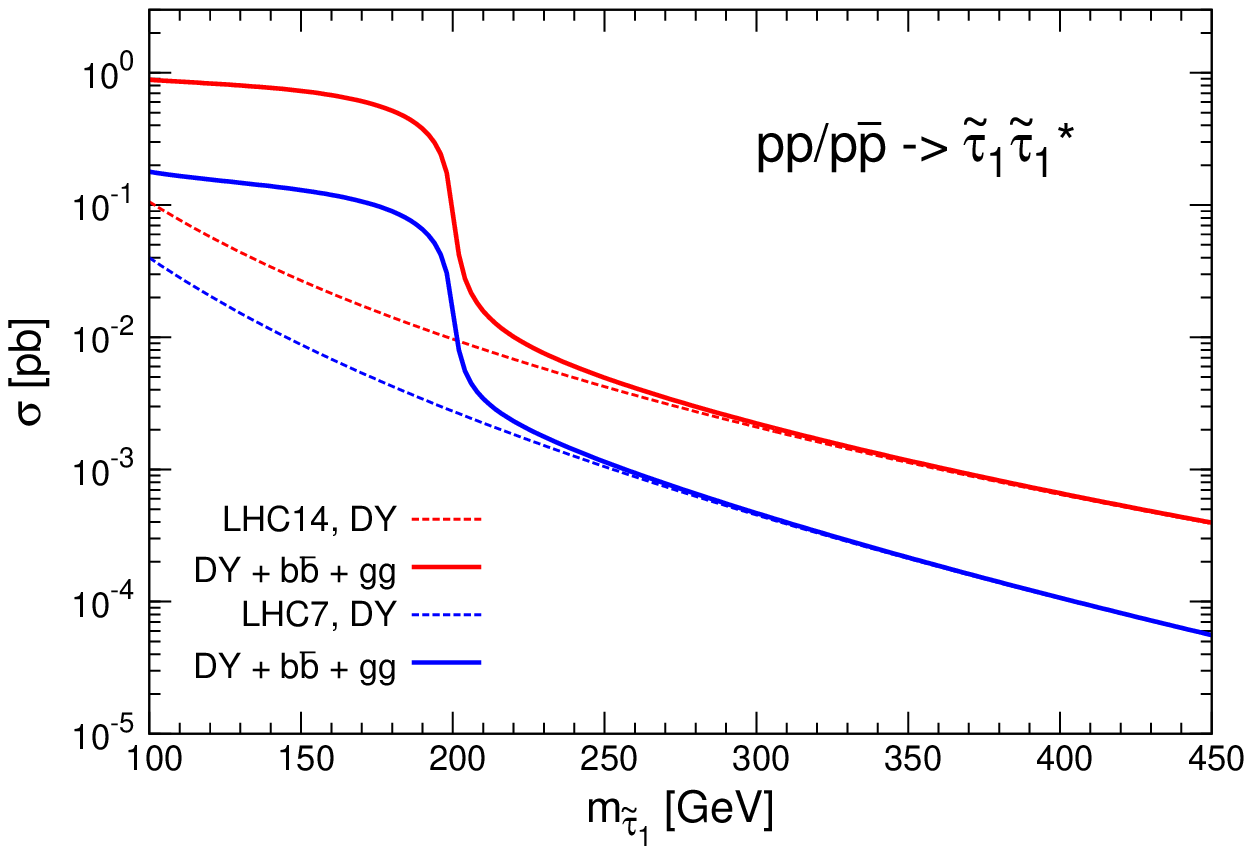}
\caption{Direct $\supertau_1$ production cross section at LHC7, LHC14 and Tevatron.
Comparison of Drell-Yan (dashed lines) and full (solid lines) cross section, including $\bbar$ and $gg$ channels.
}
\label{fig:inclusive_compare}
% scan_mstau1_tb30_compare.ps: 504x720 pixel, 72dpi, 17.78x25.40 cm, bb=0 0 504
}
% % % % % % % % % % % % % % % % % % % % % % % % 

We summarize the potential impact of the $\bbar$ and $gg$ channels 
again in \figref{fig:inclusive_compare}, where the Drell-Yan contribution (dashed lines) and the full 
cross sections (solid lines) are shown for $\staustaubar$ production at the LHC for $\SqrtS = 14~\TeV$ (top, red) 
and for $\SqrtS = 7~\TeV$ (bottom, blue). 
When going down from $14~\TeV$ to $7~\TeV$, 
the cross section decreases by up to about a factor of 5. The relative contribution of the $\bbar$ and $gg$ channels
however can be similarly important. 
Thus, for both $\SqrtS=7~\TeV$ and $14~\TeV$ (and also $\SqrtS=8~\TeV$),
the $\bbar$ and $gg$ channels should not be neglected in a precise cross section prediction.

% ____________________________________________________________________
\section{Direct production of long-lived staus within the CMSSM}
\label{sec:cmssm}
% ____________________________________________________________________

% % % % % % % % % % % % % % % % % % % % % % % % 
\FIGURE[t]{%
\includegraphics[width=.95\textwidth]{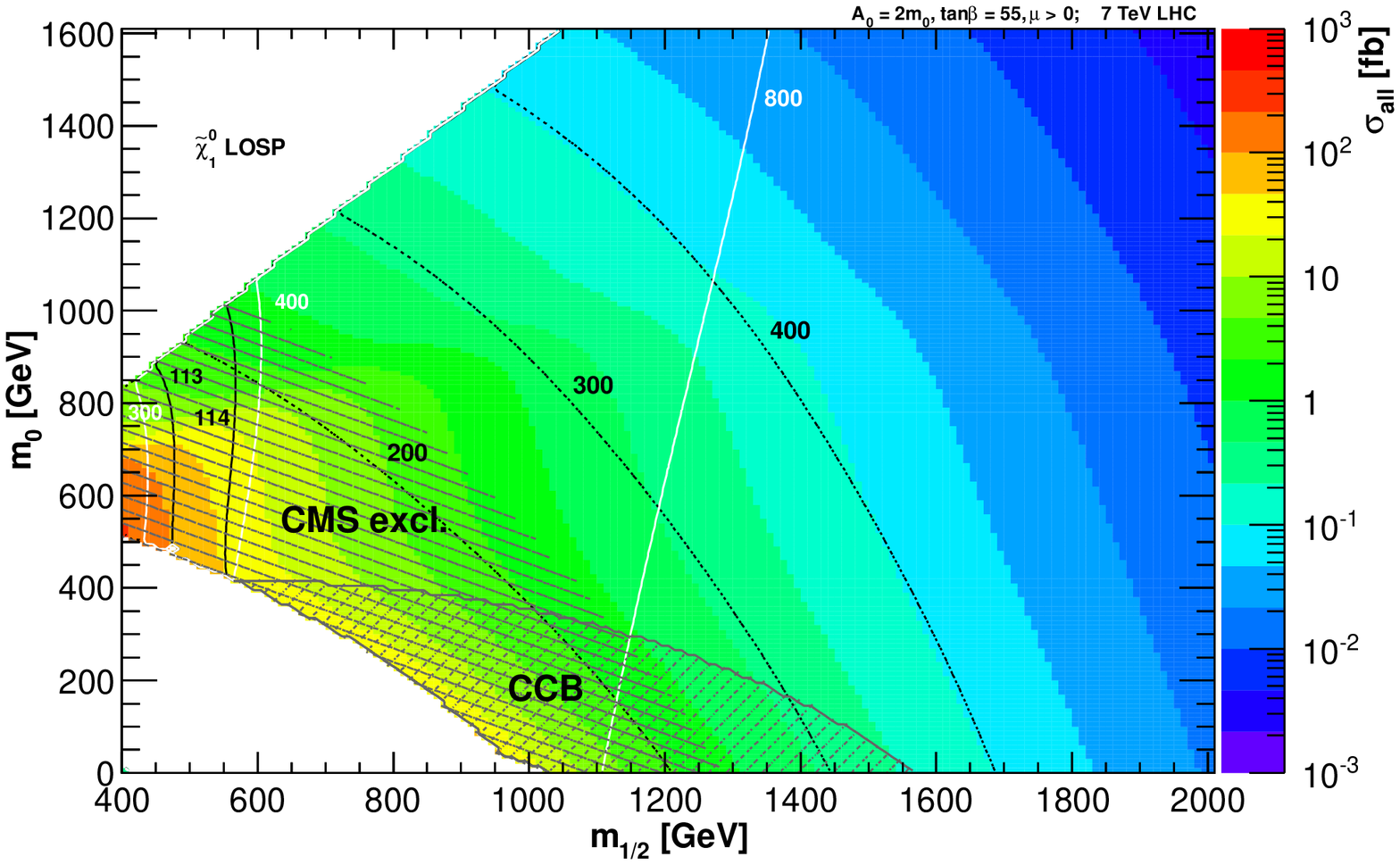}%\\%
%\includegraphics[width=.95\textwidth]{./plots/scanCMSSM14tevlabels.eps}%
 % scan_xsection.eps: 0x0 pixel, 300dpi, 0.00x0.00 cm, bb=50 50 554 770
\caption{Colored contours in the CMSSM $m_0$-$m_{1/2}$ plane with
 $\TB=55$, $A_0=2\mzero$, and $\mu>0$ of the total direct $\stau\staubar$ 
production cross section (in femtobarn) after cuts at the LHC 
with $\SqrtS = 7~\TeV$. 
The white region in the lower left is excluded because of a tachyonic spectrum,
impossible EWSB, or $\mstau \le 82~\GeV$. 
In the upper white area the lightest neutralino $\neu_{1}$ is the LOSP. 
The lower hatched area is in tension with CCB constraints. 
The solid black lines indicate the mass of the lighter Higgs boson $\hh$.
% , $\mh < 114~\GeV$.  
The dashed black lines are contours of constant \stau masses, 
% with $\mstau=100, 200, 400, 500~\GeV$,
the white lines refer to the heavier Higgs boson  $\HH$.
%  = 300, 400, 800~\GeV$. 
The hatched area at small \stau masses is excluded by recent CMS searches.  } 
\label{fig:cmssm_LHC}
}
% % % % % % % % % % % % % % % % % % % % % % % % 

Now we turn to the CMSSM as a benchmark model. We consider the \mzero-\mhalf plane 
of the CMSSM with $A_0=2\mzero$, $\TB=55$, and $\mu >0$.
Here one often finds a \stau LOSP. Moreover, this parameter choice is cosmologically 
motivated by the possibility of a long-lived \stau, due to an extremely weakly interacting dark matter
candidate such as the axino or gravitino and by exceptionally small stau yields~\cite{Pradler:2008qc}.
In this plane, \TB is large and the \stau prefers to be right-handed ($\thetastau >\pi/4$;
this is generic in the CMSSM due to the different running of the left-handed and
right-handed soft masses), 
and the $\bbar$ and $gg$ channels can give large contributions to the stau production
cross section. 
In the following, our cross section predictions include the Drell-Yan channels with NLO K-factors, 
the $\bbar$~annihilation and the $gg$~fusion contributions. Additionally we apply the following 
experimental cuts used by collider experiments to discriminate from SM backgrounds in their CHAMP
searches:
\begin{align}
\begin{split}
\label{eq:cuts}
 \pT > 40~\mbox{GeV},\qquad &0.4 < \beta < 0.9,\qquad \vert \eta \vert < 2.4\, ,
\end{split}  
\end{align}  
where $\pT$ denotes the transverse momentum, $\beta$ the velocity, and $\eta$ the pseudorapidity of the 
produced CHAMP. These cuts reduce the cross section by about $30\%$.\\

The resulting cross section is shown in \figref{fig:cmssm_LHC}. It depends mainly 
on \mstau and \mHH and varies over several orders of magnitude in the given parameter range.
At the LHC with $\SqrtS = 7~\TeV$, it reaches $10^2~\fbarn$ for $\mzero \lesssim
700~\GeV$,  
$\mhalf \lesssim 500~\GeV$ and drops to $10^{-3}~\fbarn$ for, \eg, $\mzero\sim\mhalf \sim 2~\TeV$.

In a recent analyses~\cite{recentCMS} using an integrated luminosity of 
$\mathcal{L} = 4.7\,\fbarn^{-1}$, the CMS experiment excludes long-lived staus with
\begin{align}
\label{CMSlimit}
m_{\stau} \lesssim 232~\GeV \, .
\end{align}
A thereby excluded parameter region of the considered CMSSM plane is indicated in \figref{fig:cmssm_LHC}.
Comparing with figure 12 of \cite{staupaper}, this new limit seems to disfavour the cosmologically appealing possibility of exceptionally small stau yields in this CMSSM plane (and also in the CMSSM in general).   
The limit (\ref{CMSlimit}) set by the CMS collaboration is based on the assumption of directly produced \staustaubar pairs via the \DY channel at NLO and should thus be considered as fairly model independent. 
Determination of the exact discovery reach and/or the exclusion limits including the additional
$\bbar$ and $gg$ channels should be performed in context of a detailed study including 
detector effects.

 %%%%%%%%%%%%%%%%%%%%%%%%%%%%%%%%%%%%%%%%%%%%%%%%%
\section{Conclusions}
\label{sec:conclusion}
%%%%%%%%%%%%%%%%%%%%%%%%%%%%%%%%%%%%%%%%%%%%%%%%%
We have studied the direct hadronic production of a pair of staus $\stau\stau^*$
within the MSSM. 
In addition to the well-known Drell-Yan process,
we considered production processes initiated by $\bbar$~annihilation and gluon fusion,
with all third-generation mixing effects taken into account. In parameter regions with non-negligible mixing
these contributions can enhance the direct production cross section significantly and
should always be included.
These predictions are independent of the stau lifetime and 
applicable in $\neu_1$ LSP scenarios with the $\stau$ being the next-to-LSP
as well as in settings in which the $\stau$ is long-lived.

Within the CMSSM and assuming the \stau being long-lived on the scale of 
colliders, we have provided cross section predictions 
for direct stau production at the LHC with $\SqrtS=7~\TeV$. Recent
exclusion limits from the experiments at the LHC are interpreted in this framework. 
From this, at least within the CMSSM, the possibility of exceptionally small stau yields seems
to be disfavoured. The current $\SqrtS=8~\TeV$ run of the LHC will fully probe the 
viability of such yields within the CMSSM.

Once a deviation from SM backgrounds is observed, the parameters of the underlying 
theory have to be determined. In our detailed analysis~\cite{staupaper} we offer various 
ideas how, here, direct stau production can be of utter importance. Measuring the 
enhancement due to $\bbar$~annihilation and gluon fusion together with kinematic 
distributions can even offer the possibility to test early Universe implications in the laboratory.

\end{document}